\documentclass[a4, 11pt]{article}
\begin{document}
\title{$\eta$-trigonometric states of four qubits and entanglement measures}
\author{Andrzej M. Frydryszak~$^\dag$\\
{\em $^\dag$~Institute of Theoretical Physics,University of Wroc{\l}aw,}\\
{\em Wroc{\l}aw, Poland}}
\maketitle
\abstract{Entanglement of four qubit pure states defined by the $\eta$-trigonometric functions is studied. We analyze the behavior of two recently proposed symmetric entanglement monotones on the chosen $n=4$ qubit states .}\\[3mm]


\section{Introduction}
The question of entanglement became one of the most intensively studied problems of nowadays physics \cite{QI}. The theoretical and experimental aspects of this beautiful quantum effect have to
be understood before we will be able to construct hardware as well as software of the quantum computer. Despite the immense progress in understanding there are still lots of questions to be answered. One of them is the choice of a proper entanglement monotone for multi-qubit states $n>3$, in particular for the multi-qubit pure states.
As it is well recognized, in multipartite systems, the quantification of entanglement  is highly nontrivial task. On the one hand various (SLOCC) inequivalent types of
entangled pure states of multiqubit systems for n$\leq$4 are known, but for the n=4 there is no consensus which entanglement monotone is the proper choice \cite{ckw, bren, chte-dok, ver-deh-dem, ren-zho}.

Recently, there has been proposed a new formalism that is very convenient in answering such questions \cite{mandi, mandi2, nqm}. Namely,  the generalized $\eta$-Hilbert space approach \cite{nqm}, where one introduces the space of functions of nilpotent commuting variables $\mathcal{F}(\eta_1,\dots,\eta_n)$, appropriate scalar product and necessary other objects. Within such a space it is possible to  compactly formulate criteria of separability and to construct invariants and entanglement monotones, also known from other approaches. The algebraical structure of such $\eta$-Hilbert space
allows numerous simplifications, in comparison to the conventional approach with "binary basis" notation. In particular it turns out that many interesting entangled pure states like the Werner states $\psi_W$ or the Greenberger - Horne - Zeilinger states $\psi_{GHZ}$ are elementary functions in the $\mathcal{F}(\eta_1,\dots,\eta_n)$ space \cite{nqm, ncm}. One of the sets of elementary functions in this new approach is the set of the generalized trigonometric $\eta$-functions. It is intriguing that some of the pure entangled states considered lately in the literature in the context of nontrivial entanglement of three or four qubits expressed in the $\eta$-variables fall into the trigonometric family.

In the present contribution we want to discuss the $\eta$-trigonometric functions in the case of  $n=4$ where arises interesting question of preferable entanglement monotone. We shall describe two symmetric monotones which seem to be good candidates for being the universal entanglement monotone for $n=4$. We focus on the pure state case only. In the next section we present relevant $\eta$-functions and  relate them to the combinations of conventional entangled states. Then, in the Sec. \ref{mono}, the characterization of the entanglement of these states is given, then extended to the family $G_{abcd}$ of four-qubit pure states according to Verstraete et al. \cite{ver-deh-dem-ver}.
\section{Trigonometric $\eta$-functions}
Trigonometric  $\eta$-functions are defined \cite{nqm} in analogy to the conventional case with the use of the formal series in the nilpotent commuting variables $\vec{\eta}=(\eta_1, \eta_2,\dots ,\eta_n)$  ($\eta_i^2=0$; $i=1,2,\dots, n$ moreover $\eta_i$ are indepentent i.e. $\eta_1\eta_2\dots\eta_n\neq 0$)
\begin{equation}
cos(\eta_1+\dots+\eta_n)=\sum_{k=0}(-1)^k\frac{(\eta_1+\dots+\eta_n)^{2k}}{(2k)!}
\end{equation}
\begin{equation}
sin(\eta_1+\dots+\eta_n)=\sum_{k=0}(-1)^k\frac{(\eta_1+\dots+\eta_n)^{2k+1}}{(2k+1)!}
\end{equation}
Despite nilpotent terms in the expansion the trigonometric identity is valid:  $cos^2(F) + sin^2 (F)=1$, for any $\eta$-function $F=\sum_{k=0}^n\sum_{I_k}F_{I_k}{\eta^{I_k}}$ with $\eta^{I_k}=\eta_{i_1}\eta_{i_2}\dots \eta_{i_k}$.
Let us recall that for the n=2 explicit formulas are
\begin{eqnarray}
cos(\eta^1+\eta^2)&=&1-\eta^1\eta^2\\
sin(\eta^1+\eta^2)&=&\eta^1+\eta^2
\end{eqnarray}
Relation to the binary notation of the qubit states is given in the following way e.g for two qubits: $1=|00\!\!>$, $\eta_1=|10\!\!>$, $\eta_2=|01\!\!>$ and $\eta_1\eta_2=|11\!\!>$.
We can normalize these states using $\mathcal{N}$-scalar product in the space of $\eta$-functions \cite{nqm}
\begin{equation}
<F,~G>_{\mathcal{N}}=\int F^*(\vec{\eta})G(\vec{\eta})e^{<\vec{\eta}^*,\vec{\eta}>}~d\vec{\eta}^*~d\vec{\eta}
=\int F^*(\vec{\eta})G(\vec{\eta})d\mu(\vec{\eta}^*,~\vec{\eta})
\end{equation}
where
\begin{equation}
F^*(\vec{\eta})=\sum_{k=0}^n\sum_{I_k}F^*_{I_k}{\eta^{I_k}}^*
\end{equation}
and $*$ denotes complex conjugation, $F_{I_k}$ are complex numbers. In components we have
 \begin{equation}\label{scalar}
<F,~G>_{\mathcal{N}}=\sum_{k=0}\sum_{I_k}F^*_{I_k}G_{I_k}
\end{equation}
The  one qubit states are realized in this formalism by the
$\eta$-functions of one variable.
In particular, $\eta$-scalar product of $F(\eta)$ and $G(\eta)$
functions takes simple form
\begin{equation}
<F,~G>_{\mathcal{N}}=F^*_0G_0+F^*_1G_1
\end{equation}
For our two qubit trigonometric states we get the following normalized functions
\begin{eqnarray}
\psi_{GHZ-}&=&\frac{1}{\sqrt{2}}cos(\eta_1+\eta_2)=\frac{1}{\sqrt{2}}(|00\!\!>-|11\!\!>)\\
\psi_{W}&=&\frac{1}{\sqrt{2}}sin(\eta_1+\eta_2)=\frac{1}{\sqrt{2}}(|01\!\!>+|10\!\!>)
\end{eqnarray}
The last expressions in above formulas show, how the trigonometric $\eta$-functions for two qubits read in the so called binary basis. The case of three $\eta$ variables gives also the GHZ and W -states for three qubits, but here in addition emerges the cluster Wrener state. Namely,
\begin{eqnarray}
cos(\eta^1+\eta^2+\eta^3)&=&1-\eta^1\eta^2-\eta^1\eta^3-\eta^2\eta^3\\
sin(\eta^1+\eta^2+\eta^3)&=&\eta^1+\eta^2+\eta^3-\eta^1\eta^2\eta^3
\end{eqnarray}
what after normalization gives
\begin{eqnarray}
\psi_c^{(3)}=\frac{1}{2}cos(\eta^1+\eta^2+\eta^3)&=&\frac{1}{2}
(1-\eta^1\eta^2-\eta^1\eta^3-\eta^2\eta^3)=\\\nonumber
&=&\frac{1}{2}(1-\frac{1}{\sqrt{3}}\psi_{CW})\\
\psi_s^{(3)}=\frac{1}{2}sin(\eta^1+\eta^2+\eta^3)&=&\frac{1}{2}
(\eta^1+\eta^2+\eta^3-\eta^1\eta^2\eta^3)=\\\nonumber
&=&\frac{1}{2}(1+\sqrt{3}\psi_W
-\sqrt{2}\psi_{GHZ}) \end{eqnarray}
In the binary basis above states take the following form
\begin{eqnarray}
\psi_c^{(3)}&=&\frac{1}{2}(|000\!\!>-|110\!\!>-|101\!\!>-|011\!\!>)\\
\psi_s^{(3)}&=&\frac{1}{2}(|100\!\!>+|010\!\!>+|001\!\!>-|111\!\!>)
\end{eqnarray}
and the cluster Werner state is given as $\psi_{CW}=\frac{1}{\sqrt{3}}(|110\!\!>+|101\!\!>+|011\!\!>)$.
Finally let us introduce states the most interesting for us because of their entanglement properties. Explicit form of these trigonometric $\eta$-functions is the following
\begin{eqnarray}
cos(\eta^1+\eta^2+\eta^3+\eta^4)
&=&1-\eta^1\eta^2-\eta^1\eta^3-\eta^1\eta^4-\eta^2\eta^3\\\nonumber
&-&\eta^2\eta^4-\eta^3\eta^4+\eta^1\eta^2\eta^3\eta^4\\
sin(\eta^1+\eta^2+\eta^3+\eta^4)&=&\eta^1+\eta^2+\eta^3+\eta^4
-\eta^1\eta^2\eta^3\\\nonumber
&-&\eta^1\eta^2\eta^4-\eta^1\eta^3\eta^4-\eta^2\eta^3\eta^4
\end{eqnarray}
For calculating the values of entanglement monotone we will need normalized functions, hence
\begin{eqnarray}
\psi_c^{(4)}=\frac{1}{2\sqrt{2}}(cos(\eta^1+\eta^2+\eta^3+\eta^4)
&=&\frac{1}{2}(\psi_{GHZ}-\sqrt{3}\psi_{CW})\\
\psi_s^{(4)}=\frac{1}{2\sqrt{2}}sin(\eta^1+\eta^2+\eta^3+\eta^4)
&=&\frac{1}{\sqrt{2}}(\psi_W-\star\psi_W),
\end{eqnarray}
where $\star\psi_W$ is the dual state to the Werner one. In the binary basis they are given in the form $\psi_W=\frac{1}{2}(|1000\!\!>+|0100\!\!>+|0010\!\!>+|0001\!\!>)$; $\star\psi_W=\frac{1}{2}(|1110\!\!>\!+|1101\!\!>\!+|1011\!\!>\!+|0111\!\!>)$ and
$\psi_{CW}=\frac{1}{\sqrt{6}}(|1100\!\!>\!+|1010\!\!>\!+|1001\!\!>\!+|0110\!\!>
+|0101\!\!>+|0011\!\!>)$. Then $\psi_c^{(4)}$ and $\psi_s^{(4)}$
\begin{eqnarray}
\psi_c^{(4)}&=&\frac{1}{2\sqrt{2}}(|0000\!\!>-|1100\!\!>-|1010\!\!>-|1001\!\!>
-|0110\!\!>\\\nonumber
&-&|0101\!\!>-|0011\!\!>+|1111\!\!>)\\
\psi_s^{(4)}&=&\frac{1}{2\sqrt{2}}(|1000\!\!>+|0100\!\!>+|0010\!\!>+|0001\!\!>
-|1110\!\!>\\\nonumber
&-&|1101\!\!>-|1011\!\!>-|0111\!\!>)
\end{eqnarray}
In addition to the above states let us consider also known in the literature the cluster type state of the form
\begin{equation}
\psi_{cs}=\frac{1}{2}(|0000\!\!>+|0011\!\!>+|1100\!\!>-|1111\!\!>),
\end{equation}
what in terms of the $\eta$-functions reads as
\begin{equation}\label{cs}
\psi_{cs}=\frac{1}{2}(cos(\eta_1\eta_2+\eta_3\eta_4)+sin(\eta_1\eta_2+\eta_3\eta_4)).
\end{equation}
In the following section we shall test behavior of two entanglement monotones using introduced above set of states.
\section{Entanglement of the n=4 trigonometric pure states}\label{mono}
The entanglement monotones for the $n=2$ and $n=3$ are well studied. For the first case situation is very comfortable and entanglement can be effectively characterized as for the pure states as for mixed states. There always exists the Schmidt decomposition and various options in equivalent definitions of entanglement monotone - the concurrence. For the $n=3$ description of entanglement is complicated by the fact that there is no literal  Schmidt decomposition and there are only some generalizations with somehow weaker properties then for $n=2$.
However it is still relatively simple to define proper entanglement measure, but appears issue of the distinction between full separability and partial separability of states and that is why relative concurrences and the three tangle are used. From the point of view of the SLOCC classification there are only two inequivalent entanglement classes, one is represented by the GHZ-state and the other one by the W-state.
The $n=4$ case is more complicated, but much is already known what we can't say about $n\geq 5$ multi-partite entanglement description. Here, form the SLOCC classification one gets nine inequivalent entangled states. The approach based on the polynomial invariants is more complicated but amenable. The set of necessary invariants can be found using combs.
It seems that the symmetric entanglement monotones (permutation invariant) are
the
best choice for characterization of the multipartite entanglement. The way qubits
are labeled should not influence its value. For the four qubits natural candidates for the symmetric entanglement monotones are two invariants $\mathcal{F}'_2$ \cite{ren-zho} and $\mathcal{F}_3$ based on the ones introduced by Osterloh and Siewert \cite{ost-sie, ren-zho}. They can be expressed in terms of the Schl\"{a}fli basis $\{\mathcal{H}, W,\Sigma, \Pi\}$, where $W=D_{xy}+D_{xz}+D_{xt}$, $\Sigma=L^2+M^2+N^2$ and $\Pi=(L-M)(M-N)(N-L)$. Explicitly they have the following form
\begin{eqnarray}
|\mathcal{F}_3|&=&32|\mathcal{H}^6-24\mathcal{H}^2\Sigma-64\Pi|\\
|\mathcal{F}_2'|&=&16|3\mathcal{H}^4-16\mathcal{H}W+8\Sigma|
\end{eqnarray}
Let us evaluate $|\mathcal{F}_3|$ and $|\mathcal{F}_2'|$ on the $n=4$ $\eta$-trigonometric states. It turns out that for both cases $W$, $\Sigma$ and $\Pi$ vanishes and the only contribution comes from the $W$ invariant. We have
\begin{eqnarray}
|\mathcal{F}_3(\psi_c)|&=&|\mathcal{F}_3(\psi_s)|=\frac{1}{2}\\
|\mathcal{F}_2'(\psi_c)|&=&|\mathcal{F}_2'(\psi_s)|=3
\end{eqnarray}
It is interesting that for
$\psi_c(\alpha)\equiv sin\alpha\, \psi_{CW}+cos\alpha\,\psi_{GHZ}$ we get constant value of above entanglement measures while for
$\psi_s(\alpha)\equiv cos\alpha \,\psi_W+\sin\alpha\,\star\psi_W$
both entanglement monotones vary between zero and its maximal value for the $\psi_s(\alpha)$ family i.e. $|\mathcal{F}_3(\psi_s(\alpha))|=\frac{1}{2}\sin^6(2\alpha))$ and $|\mathcal{F}_2'(\psi_s)|=3\sin^4(2\alpha)$. As we have mentioned before the entanglement of the $n=4$ Werner state and its dual state is not detected by both entanglement monotones.
Let us look, in more detail, at the case of the $\psi_c(\alpha)$. These states belong to the family  $G_{abcd}$ according to the classification given in \cite{ver-deh-dem-ver} (nine families classified by SLOCC transformations). General $\eta$-function representing such a state can be written as follows
\begin{eqnarray}
\psi_{abcd}&=&\frac{a+d}{2}e^{\vec{\eta}}+\frac{a-d}{2}(\cos(\eta_1-\eta_2)-
\cos(\eta_3+\eta_4))\\\nonumber
&&+\frac{b+c}{2}(\cos(\eta_1-\eta_3)-\cos(\eta_2+\eta_4))\\\nonumber
&&+\frac{b-c}{2}(\cos(\eta_1-\eta_4)-\cos(\eta_2+\eta_3))
\end{eqnarray}
We obtain the $\psi_{c}(\alpha)$ by taking $a=\cos\alpha+\sin\alpha$, $b=\sin\alpha$, $c=0$ and $d=\cos\alpha-\sin\alpha$. As it is known two states from the one of the nine families found in Ref.\cite{ver-deh-dem-ver} may be in the same orbit, that is why  the $|\mathcal{F}_3|$ and $|\mathcal{F}'_2|$ are constant for $\psi_c(\alpha)$. However it is not generic situation. To see this, let us move along $\psi_{abcd}$ family, taking parameters such that $a=b$ and $c=d$. Denoting
\begin{equation}\label{zeta}
\zeta=\left(\frac{2ad}{a^2+d^2}\right)^2
\end{equation}
we have that
\begin{equation}
|\mathcal{F}_3(\psi_{ad})|=|\frac{1}{2}-\frac{3}{2}\zeta^4+\zeta^6|
=|(\zeta-1)(\zeta+\frac{1}{2})|
\end{equation}
and
\begin{equation}
|\mathcal{F}'_2(\psi_{ad})|=|3-2^2\zeta+\zeta^2|
=|(\zeta-1)(\zeta-3)|
\end{equation}
Again, when one of the parameters $a$ or $d$ vanishes, we obtain invariant $|\mathcal{F}_3(\psi_{ad})|$ and $|\mathcal{F}'_2(\psi_{ad})|$, but otherwise
these entanglement monotones vary for states from the $G_{abcd}$. In particular
we find states for which above entanglement measures simultaneously vanish (taking into account (\ref{zeta}) we see that both polynomials have only one, common zero). Hence the entanglement monotones $|\mathcal{F}_3|$, $|\mathcal{F}'_2|$ indicate that the states
\begin{equation}
\psi_a=a(e^{\vec{\eta}}+(\cos(\eta_1-\eta_3)-\cos(\eta_2+\eta_4)))
\end{equation}
and
\begin{equation}
\psi_d=d(\cos(\eta_1-\eta_2)-\cos(\eta_3+\eta_4)
+\cos(\eta_1-\eta_4)-\cos(\eta_2+\eta_3))
\end{equation}
do not exhibit genuine four-qubit entanglement. Indeed, $\psi_a\sim\psi_W^{(13)}\psi_W^{(24)}$ and $\psi_d\sim\psi_{GHZ}^{(13)}\psi_{GHZ}^{(24)}$. So, there is only residual two-qubit entanglement.

Finally let us consider the cluster type state $\psi_{cs}$
given by the Eq.(\ref{cs}) . Here situation is different, the Cayley determinant $\mathcal{H}(\psi_{cs})$ vanishes as well as the $W(\psi_{cs})$, but $\Pi(\psi_{cs})=(\frac{1}{2})^{11}$ and
$\Sigma=(\frac{1}{2})^{7}$. This means that
\begin{equation}
|\mathcal{F}_3(\psi_{cs})|=|\mathcal{F}_2'(\psi_{cs})|=1
\end{equation}
The question if the considered monotones for the system of $n=4$ qubits qualify to be proper genuine entanglement measures is still open, but as present analysis shows their symmetry properties and behavior on $\eta$-trigonometric states gives an indication that they are a reasonable candidates.
%
%
%

\end{document}